%% file: iwara2013_Contrera_et_al_final.tex
\documentclass[12pt]{article}
\usepackage{amssymb}
\usepackage{float}
\usepackage[dvipsnames]{xcolor}

\input econfmacros-iwara13.tex
\def\Title#1{\begin{center} {\Large {\bf #1} } \end{center}}

\begin{document}

\Title{Hadron-Quark Phase Transition in Quark-Hybrid Stars}



\begin{raggedright}
{Gustavo A. Contrera\index{Contrera, G. A.} \footnote{Home address: IFLP, CONICET - Dpto. de F{\'i}sica, UNLP, La Plata, Argentina;\\ Gravitation, Astrophysics and Cosmology Group, Facultad de Ciencias Astron{\'o}micas y Geof{\'i}sicas,\\ Paseo del Bosque S/N (1900), Universidad Nacional de La Plata UNLP, La Plata, Argentina}\\
\it  Department of Physics, San Diego State
University, San Diego, CA 92182, USA\\
{\tt Email: guscontrera@gmail.com}}\\
{William Spinella\index{Spinella, W.}\\
\it Computational Science Research Center \& Department of Physics, San Diego State University, San Diego, CA 92182, USA\\
{\tt Email: spinella@rohan.sdsu.edu}}\\
{Milva Orsaria\index{Orsaria, M.}\\
\it CONICET, Rivadavia 1917, 1033 Buenos Aires, Argentina; Gravitation, Astrophysics and Cosmology Group, Facultad de Ciencias Astron{\'o}micas y Geof{\'i}sicas, Paseo del Bosque S/N (1900), Universidad Nacional de La Plata UNLP, La Plata, Argentina\\
{\tt Email: morsaria@fcaglp.unlp.edu.ar}}\\
{Fridolin Weber\index{Weber, F.}\\
\it Department of Physics, San Diego State
University, San Diego, California 92182, USA \& Center for
Astrophysics and Space Sciences, University of California, San Diego,
La Jolla, California 92093, USA \\
{\tt Email: fweber@mail.sdsu.edu}}
\bigskip\bigskip
\end{raggedright}

\section{Introduction}

The recent discovery of the two-solar mass neutron stars $J1614-2230$
($1.97 \pm 0.04 M_{\odot}$)~\cite{Demorest2010} and $J0348+0432$
($2.01 \pm 0.04 M_{\odot}$)~\cite{Antoniadis13} allows us to consider the
possible existence of deconfined quarks in the cores of neutron
stars~\cite{Orsaria:2013, Orsaria:2014}. Nevertheless if the dense interior
of a neutron star is indeed converted to quark matter, it must be
three-flavor quark matter since it has lower energy than two-flavor
quark matter. And just as for the hyperon content of neutron stars,
strangeness is not conserved on macroscopic time
scales, which allows neutron stars to convert confined hadronic matter
to three-flavor quark matter until equilibrium brings this process to
a halt.  As first realized by Glendenning~\cite{glendenning00}, the
presence of quark matter in neutron stars enables the hadronic regions
of the mixed phase to become more isospin symmetric than in the pure
phase by transferring electric charge to the quark phase. The symmetry
energy can be lowered thereby at only a small cost in rearranging the
quark Fermi surfaces. The stellar implication of this charge rearrangement is
that the mixed phase region of a neutron star will have positively
charged regions of nuclear matter and negatively charged regions of
quark matter~\cite{glendenning00}. This should
have important implications for the electric and thermal properties of
neutron stars.  First studies of the transport properties of
quark-hybrid neutron star matter have been reported in
~\cite{reddy00:a,na12:a}.

It has been shown
\cite{Alford:2001zr,Voskresensky:2002hu,Tatsumi:2002dq,Endo:2011em}
that the appearance of a mixed phase of quarks and hadrons in NSs depends
on the surface tension between nuclear and quark matter, which should be
around $5-30$ MeV/fm$^3$. Although recent studies about the nucleation process during the phase transition
predict the value of the surface tension, their results vary and are strongly model dependent \cite{Palhares:2010,Pinto:2012,Ke:2013wga,Carmo:2013fr,Lugones:2013ema}. Thus, the discussion concerning the appearance of a mixed phase in NSs remains open.

Our study is based on NSs containing deconfined quark matter, i.e.\ quark-hybrid stars (QHSs). To describe the quark matter phase, we use a non-local extension of the SU(3) Nambu Jona-Lasinio (NJL) model with vector interactions, whereas to represent the hadronic phase we consider a non-linear Walecka model using parametrization NL3~\cite{Lalazissis}. A phase transition between these two phases can be constructed via the Gibbs conditions, imposing global electric charge neutrality and baryon number conservation. We find that the non-local NJL model predicts the existence of extended regions of mixed quark-hadron (quark-hybrid) matter in high-mass neutron stars with masses of 2.0 to $2.4\, M_{\odot}$.

\section{Modeling of the Mixed Phase}

To determine the mixed phase region of quarks and hadrons, we start
from the Gibbs condition for pressure equilibrium between confined
hadronic ($P^H$) matter and deconfined quark ($P^q$) matter.  The
Gibbs condition is given by
\cite{glendenning00}
\begin{eqnarray}
 P^H(\mu_b^H, \mu_e^H, \{\phi \} ) = P^q(\mu_b^q, \mu_e^q, \{\psi \} )
 \, ,
\label{eq:GibbsP}
\end{eqnarray}
with $\mu_b^H = \mu_b^q$ for the baryon chemical potentials and
$\mu_e^H=\mu_e^q$ for the electron chemical potentials in the hadronic
($H$) and quark ($q$) phase, respectively. By definition, the quark
chemical potential is given by $\mu_b^q=\mu_n/3$, where $\mu_n$ is the
chemical potential of the neutron. The quantities $\{ \phi \}$ and $\{
\psi\}$ in Eq.\ (\ref{eq:GibbsP}) stand collectively for the field
variables and Fermi momenta that characterize the solutions to the
equations of confined hadronic matter and deconfined quark matter,
respectively.  In the mixed phase, the baryon number density, $n_b$,
and the energy density, $\varepsilon$, are given by
\cite{glendenning00}
\begin{equation}
    n_b = (1-\chi) n_b^H + \chi n_b^q \, ,
\end{equation}
and
\begin{equation}
    \varepsilon = (1-\chi) \varepsilon^H + \chi\varepsilon^q\, ,
\end{equation}
where $n_b^H$ ($\varepsilon^H$) and $n_b^q$ ($\varepsilon^q$) denote
the baryon number (energy) densities of the hadron and quark phase,
respectively. The quantity $\chi \equiv V_q/V$ denotes the volume
proportion of quark matter, $V_q$, in the unknown volume $V$. By
definition $\chi$ varies between 0 and 1 depending on how
much confined hadronic matter has been converted to quark matter
\cite{glendenning00}. In
addition to the Gibbs condition (\ref{eq:GibbsP}) for pressure, the
conditions of global baryon number conservation and global electric charge
neutrality need to be imposed on the field equations.
\begin{eqnarray}
  \rho_b = \chi \, \rho_Q(\mu_n, \mu_e ) + (1-\chi) \, \rho_H (\mu_n,
  \mu_e, \{ \phi \}) \, ,
\label{eq:mixed_rho}
\end{eqnarray}
where  $\rho_Q$ and $\rho_H$ denote the baryon number densities of the
quark phase and hadronic phase, respectively. The condition of global
electric charge neutrality is given by
\begin{equation}
(1 - \chi) \sum_{i=B,l} q_i^H \, n_i^H + \chi \sum_{i=q,l}q_i^q \,
  n_i^q = 0 \, ,
\end{equation}
where $q_i$ is the electric charge of the $i$-th specie in units of
the electron charge.

For the quark sector, within the non-local NJL model, the mean-field thermodynamic potential at zero temperature is \cite{Orsaria:2014}
\begin{eqnarray}
&&\Omega^{NL} (M_f,0,\mu_f) = -\frac{N_c}{\pi^3}\sum_{f=u,d,s}
\int^{\infty}_{0} dp_0 \int^{\infty}_{0}\, dp \,\mbox{ ln }\left\{
\left[\widehat \omega_f^2 + M_{f}^2(\omega_f^2)\right]
\frac{1}{\omega_f^2 + m_{f}^2}\right\} \nonumber \\ & & -
\frac{N_c}{\pi^2} \sum_{f=u,d,s} \int^{\sqrt{\mu_f^2-m_{f}^2}}_{0}
dp\,\, p^2\,\, \left[(\mu_f-E_f) \theta(\mu_f-m_f) \right]
\\ & &
- \frac{1}{2}\left[\sum_{f=u,d,s} (\bar \sigma_f \ \bar S_f +
  \frac{G_S}{2} \ \bar S_f^2) + \frac{H}{2} \bar S_u\ \bar S_d\ \bar
  S_s\right]
- \sum_{f=u,d,s}\frac{\varpi_f^2}{4 G_V} \, , \nonumber
\label{omzerot}
\end{eqnarray}
where $N_c=3$, $E_{f} = \sqrt{p^{2} + m_{f}^{2}}$, and $\omega_f^2 =
(\,p_0\,+\,i\,\mu_f\,)^2\,+\,p^2$.  The constituent quark masses
$M_{f}$ are treated as momentum-dependent quantities and are given by
\begin{equation}
M_{f}(\omega_{f}^2) \ = \ m_f + \bar\sigma_f g(\omega_{f}^2)\, ,
\end{equation}
where $g(\omega^2_f)$ is the form factor, which we take to be Gaussian $g(\omega^2_f) = \exp{\left(-\omega^2_f/\Lambda^2\right)}$.

The inclusion of vector interactions shifts the quark chemical potential as
\begin{equation}
\mu_f\, \rightarrow \widehat{\mu}_f=\mu_f - g(\omega^2_f)\varpi_f\, ,
\label{eq:mu_f}
\end{equation}
where $\varpi_f$ represents the vector mean fields related to the vector current interaction. The inclusion of the form factor in Eq.\ (\ref{eq:mu_f}) is a particular feature of the non-local model, which renders the shifted
chemical potential momentum dependent.  Accordingly, the four momenta $\omega_f$ in the dressed part of the thermodynamic potential are modified as
\begin{equation}
\omega_f^2\,\rightarrow \widehat{\omega}_f^2 = (\,p_0\, +
\,i\,\widehat{\mu}_f\,)^2\, + \,p^2 \, .
\end{equation}
Note that the quark chemical potential shift does
not affect the non-local form factor $g(\omega_{f}^2)$, as discussed
in \cite{Dumm,Weise2011,Contrera:2012wj}, avoiding a recursive
problem. In this work we use for the NJL model the parameters listed in Refs.
\cite{Orsaria:2013,Orsaria:2014}.

 Within the stationary phase approximation, the mean-field values of
 the auxiliary fields $\bar S_f$ turn out to be related to the
 mean-field values of the scalar fields $\bar \sigma_f$
 \cite{Scarpettini}. They are given by
\begin{equation}
\bar S_f = -\, 16\,N_c\, \int^{\infty}_{0}\,dp_0 \int^{\infty}_{0}
\frac{dp}{(2\pi)^3} \, g(\omega_f^2)\, \frac{
  M_{f}(\omega_f^2)}{\widehat{\omega}^2 + M_{f}^2(\omega_f^2)}\, .
\end{equation}

Due to the charge neutrality constraint, for the quark phase we consider the three mean-field flavors $\bar \sigma_u$, $\bar \sigma_d$ and $\bar \sigma_s$, which can be obtained by solving the ``gap" equations given by \cite{Scarpettini}

\begin{eqnarray}
\bar \sigma_u + G_S\,\bar
S_u + \frac{H}{2} \, \bar S_d \bar S_s &=& 0\, , \nonumber \\
\bar \sigma_d + G_S\,\bar
S_d + \frac{H}{2} \, \bar S_u \bar S_s &=& 0\, , \\
 \bar \sigma_s + G_S\,\bar
S_s + \frac{H}{2} \, \bar S_u \bar S_d &=& 0\, , \nonumber
\end{eqnarray}
and $\varpi_{f}$ are obtained via minimizing the thermodynamic potential, $\frac{\partial \Omega^{\rm NL}}{\partial {\varpi_{f}}}= 0$.

For the hadronic phase we have used the same model and parameters detailed in
Refs. \cite{Orsaria:2013,Orsaria:2014}, also considering for this work universal coupling constants.

\begin{figure}[htb]
\begin{center}
\includegraphics[width=10.0cm]{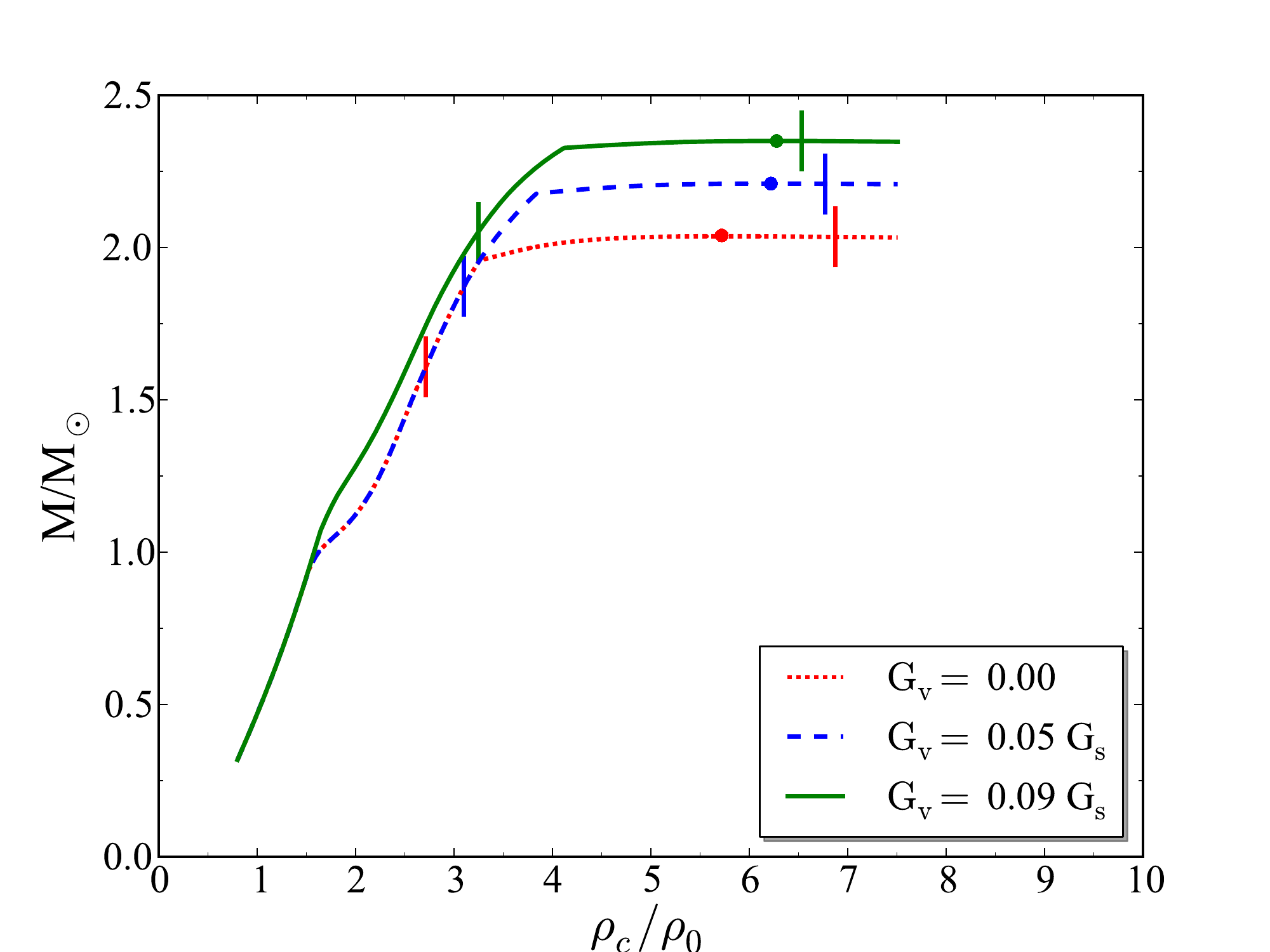}\caption{Mass versus central density of QHSs computed for the EoS constructed through  the non-linear Walecka and the non-local NJL models. Vertical bars indicate the beginning and the end of the mixed phase. Dots refer to the maximum mass star for different values of the vector interaction.}
\label{fig:mass_den}
\end{center}
\end{figure}
\begin{figure}[htb]
\begin{center}
\includegraphics[width=10.0cm]{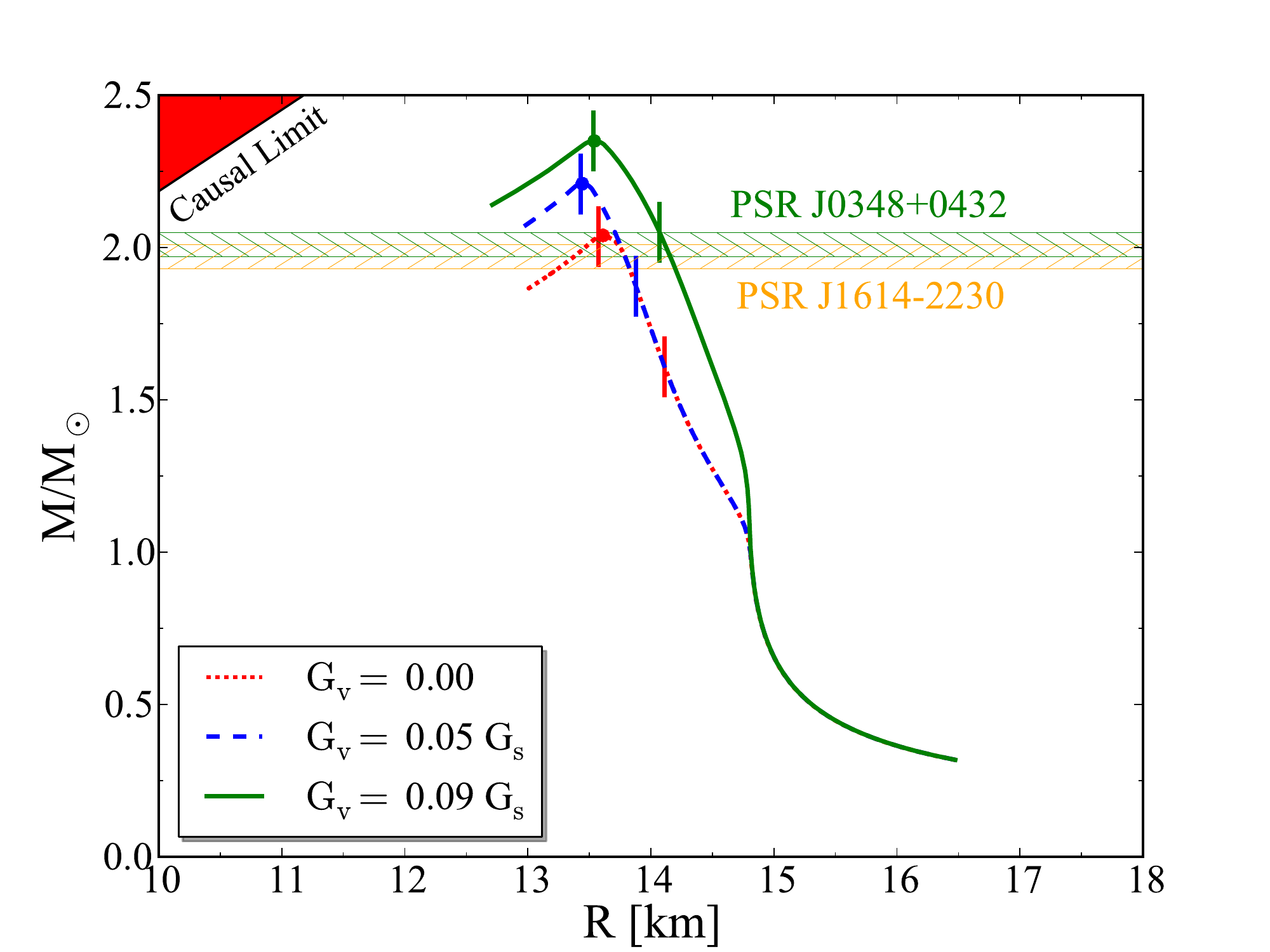}\caption{Mass-Radius relationship of QHSs computed for the EoS constructed for the non-linear Walecka and the non-local NJL models.}
\label{fig:mass_R}
\end{center}
\end{figure}

\section{Results and Conclusions}

Depending on the strength of quark vector repulsion, we find that an extended region made of a mixed phase of quarks and hadrons may exist in high-mass neutron stars with masses up to $2.0-2.4 M_{\odot}$ as can be seen in Fig~\ref{fig:mass_den}. The radii of these objects are between 12 and 13 km, as expected for neutron stars. Table~\ref{tab:quark_fraction} lists the strength of the vector interaction, the maximum mass stars, the corresponding fraction of quark matter, and the range of the mixed phase.

\begin{table}[H]
\begin{center}
\begin{tabular}{|c|c|c|c|}\hline
$G_V/G_S$ & $M_{\rm{Max}}$ ($M_{\odot}$) &  $\chi$ at $M_{\rm{Max}}$ & Mixed phase ($\rho_0$)\\ \hline
$0.00$ & $2.04$ & $0.72$ & $2.71 - 6.87$\\
$0.05$ & $2.21$ & $0.85$ & $3.10 - 6.77$\\
$0.09$ & $2.35$ & $0.92$ & $3.25 - 6.53$\\ \hline
\end{tabular}
\caption{Different strengths of the vector interaction ($G_V$) for the quark phase, maximum mass star ($M_{\rm{Max}}$), the corresponding fraction of quark matter at {$M_{\rm{Max}}$} ($\chi$), and the range of the mixed phase.}
\label{tab:quark_fraction}
\end{center}
\end{table}


For the non-local NJL model and the NL3 parametrization for the hadronic phase, we find that pure quark matter could not exist in stable neutron stars.  Only neutron stars that lie on the left of the mass peak are dense enough to contain quark matter. However, these stars are unstable against radial oscillations and thus could not exist stably in the universe. According to what is shown on Fig. \ref{fig:mass_R}, with increasing
stellar mass, all the stellar cores are composed of either nucleons, nucleons and hyperons, or a mixed phase of nucleons, hyperons, and quark matter.

\bigskip
\section*{Acknowledgments}
G.C. and M.O. thank CONICET (Argentina) for financial support and are thankful for hospitality extended to them at the San Diego State University, where much of this work was performed. F.W. acknowledges supported by the National Science Foundation (USA) under Grant PHY-0854699. This work has been partially supported by a CONICET-NSF joint project.

\end{document}

%% file: econfmacros-iwara13.tex
\def\beq{\begin{equation}}
\def\eeq#1{\label{#1}\end{equation}}
\def\eeqn{\end{equation}}

\def\beqa{\begin{eqnarray}}
\def\eeqa#1{\label{#1}\end{eqnarray}}
\def\eeqan{\end{eqnarray}}

\let\bar=\overbar

\def\Dslash{\not{\hbox{\kern-4pt $D$}}}
\def\dslash{\not{\hbox{\kern-2pt $\del$}}}

\def\msb{{\bar{\ssstyle M \kern -1pt S}}}

\usepackage{fancyhdr,graphicx}

%% file: iwara2013_Contrera_et_al_final.bbl
\begin{thebibliography}{99}


\bibitem{Demorest2010} P. B. Demorest, T. Pennucci, S. M. Ranson,
  M. S. E. Roberts and J. W. T. Hessels, Nature {\bf 467}, 1081 (2010).

\bibitem{Antoniadis13} Lynch {\it et al.} Astrophys. J {\bf
    763}, 81 (2013); J. Antoniadis {\it et al.}, Science {\bf 340},
  no.\ 6131 (2013).
\bibitem{Orsaria:2013} M. Orsaria, H. Rodrigues, F. Weber and
  G. A. Contrera, Phys. Rev. D {\bf 87}, 023001 (2013).

\bibitem{Orsaria:2014} M. Orsaria, H. Rodrigues, F. Weber and
  G. A. Contrera, Phys. Rev. C {\bf 89}, 015806 (2014).

\bibitem{glendenning00} N. K. Glendenning, {\it Compact Stars,
  Nuclear Physics, Particle Physics, and General Relativity}, 2nd
  ed. (Springer-Verlag, New York, 2000); N. K. Glendenning, Phys.\ Rev.\ D {\bf 46}, 1274 (1992); N. K. Glendenning, Phys.\ Rep.\ {\bf 342}, 393 (2001).

\bibitem{reddy00:a}
S. Reddy, G. Bertsch and M. Prakash, Phys.\ Lett.\ B {\bf 475}, 1 (2000).

\bibitem{na12:a}
X. Na, R. Xu, F. Weber and R. Negreiros, Phys.\ Rev.\ D {\bf 86}, 123016 (2012).

\bibitem{Alford:2001zr}
  M.~G.~Alford, K.~Rajagopal, S.~Reddy and F.~Wilczek,
  Phys.\ Rev.\ D {\bf 64}, 074017 (2001).

\bibitem{Voskresensky:2002hu}
  D.~N.~Voskresensky, M.~Yasuhira and T.~Tatsumi,
  Nucl.\ Phys.\ A {\bf 723}, 291 (2003).

\bibitem{Tatsumi:2002dq}
  T.~Tatsumi, M.~Yasuhira and D.~N.~Voskresensky,
  Nucl.\ Phys.\ A {\bf 718}, 359 (2003).

\bibitem{Endo:2011em}
  T.~Endo,
  Phys.\ Rev.\ C {\bf 83}, 068801 (2011); T.~Endo, arXiv:1310.0913 [astro-ph.HE].

\bibitem{Palhares:2010} L. F. Palhares and E. S. Fraga, Phys.\ Rev.\ D {\bf 82}, 125018 (2010).

\bibitem{Pinto:2012} M. B. Pinto, V. Koch and J. Randrup, Phys. Rev. C {\bf 86}, 025203 (2012).

\bibitem{Ke:2013wga}
  W.~Ke and Y.~Liu,
  arXiv:1312.2295 [hep-ph].

\bibitem{Carmo:2013fr}
  T.~A.~S.~d.~Carmo, G.~Lugones and A.~G.~Grunfeld,
  J.\ Phys.\ G {\bf 40}, 035201 (2013).

\bibitem{Lugones:2013ema}
  G.~Lugones, A.~G.~Grunfeld and M.~Al Ajmi,
  Phys.\ Rev.\ C {\bf 88}, 045803 (2013).

\bibitem{Lalazissis} G. A. Lalazissis, J. Konig and P. Ring,
  Phys.\ Rev.\ C {\bf 55}, 540 (1997).

\bibitem{Dumm} D. B. Blaschke, D. Gomez Dumm, A. G. Grunfeld,
  T. Kl\"{a}hn, and N. N. Scoccola, Phys. \ Rev.\ C {\bf 75}, 065804
  (2007).

\bibitem{Weise2011} K. Kashiwa, T. Hell, and W. Weise, Phys. Rev. D
  \ {\bf 84}, 056010 (2011).

\bibitem{Contrera:2012wj} G.~A.~Contrera, A.~G.~Grunfeld, and
  D.~B.~Blaschke, arXiv:1207.4890 [hep-ph].

\bibitem{Scarpettini} A.~Scarpettini, D.~Gomez Dumm, and
  N.~N.~Scoccola, Phys.\ Rev.\ D {\bf 69}, 114018 (2004).

\end{thebibliography}
